\documentclass{elsart}

\usepackage{amssymb}
\usepackage{epsf,colordvi,color,amsbsy}
\usepackage[dvips]{graphicx}
\begin{document}

\begin{frontmatter}

\title{On the sixty-year periodicity in climate and astronomical series}

\author[]{Stefano Sello \corauthref{}}

\corauth[]{stefano.sello@enel.it}

\address{Mathematical and Physical Models, Enel Research, Pisa - Italy}

\begin{abstract}
In a recent article by Scafetta, 2010, the author investigates whether or not the decadal and multi-decadal climate oscillations have an astronomical origin. In particular, the author note
that several global surface temperature records, since 1850, and records deduced from the orbits of the planets present very similar power spectra.
Among the detected frequencies, large climate oscillations of about 20 and 60 years, respectively, appear
synchronized to the orbital periods of Jupiter and Saturn. Other investigators have already noted that many climate, geophysical and astromomical data clearly show the appearance of a significant, approximately 60-year cycle. Of course, this cycle length is not exactly 60 years and varies by a few years (frequency band) between various climatic and astronomical phenomena.
The main aim of the present research note is to further investigate the above results, considering different long-term time series and using a proper continuous wavelet analysis. In particular, we specifically consider the feature and importance of the sixty-year periodicity, in order to
better build reliable models for climate predictions.  
\end{abstract}
\end{frontmatter}

\section{Introduction}
We recall here some brief introductive considerations of the argument under study, as well reported in previous articles and research notes. The interested reader is referred to the bibliography for further insights.
In a recent article by Scafetta, 2010, the author investigates whether or not the decadal and multi-decadal climate oscillations have an astronomical origin. It is well known that Milankovic, 1941, theorized that variations in eccentricity,
axial tilt, and precession of the orbit of the Earth determine climate features such as the 100,000 year ice age cycles of the
Quaternary glaciation over the last few million years. Moreover, the variation of the orbital parameters of the Earth is due to the gravitational
perturbations induced by the other planets of the solar system, mainly the giant ones, Jupiter and Saturn. Over a much longer
time scale, the cosmic-ray flux record well correlates with the warm and ice periods of the Phanerozoic during the last 600
million years: the cosmic-ray flux modulations are likely due to the changing galactic environment of the solar system as it
crosses the spiral arms of the Milky Way, and on shorter time scales on the solar cycle activity, (see Scafetta, 2010 and references herein).
As noted by Scafetta: "The above results suggest that the dominant drivers of the climate oscillations have a celestial origin. Therefore, it is legitimate
to investigate whether the climate oscillations, with a time scale between 1 and 100 years, can be interpreted in astronomical
terms too."  Moreover: "... several records indicate that the climate is characterized by a large quasi-60 year periodicity, plus larger secular climatic
cycles and smaller decadal cycles. All these cycles cannot be explained with anthropogenic emissions. Patterson et al. (2004) found 60-62 year cycles in sediments
and cosmogenic nuclide records in the NE Pacific. Komitov, 2009, found similar cycles in the number of the middle latitude
auroras from 1700 to 1900. A cycle of about 60 years has been detected in the number of historically recorded meteorite falls
in China from AD 619 to 1943 and in the number of witnessed falls in the world from 1800 to 1974 (Yu et al., 1983). Ogurtsov
et al. (2002) found a 60-64 year cycle in 10Be, 14C and Wolf number over the past 1000 years. The existence of a 60-year
signal has been found in the Earth angular velocity and in the geomagnetic field (Roberts et al., 2007). These results clearly
suggest an astronomical origin of the 60-year variability found in several climatic records." The main idea of the above author is that:
"the climate oscillations are described by a given, even if still unknown, physical function that depends on the orbits of the planets and their positions."
In particular, "a set of planetary frequencies can be determined by using as proxy the distance of the Sun about the center of mass of
the solar system (CMSS), or it is possible to choose its speed (SCMSS)."
Other investigators, have examined the appearance of the approximately 60-year cycle that shows up in many areas, noting that this cycle length is not exactly 60 years and varies by a few years between various climatic phenomena and locations.
In particular, in their reports are well described the multidecadal periodicities coming from: the Global Temperature Anomalies, the Atlantic Multidecadal Oscillation (AMO); the 
Pacific Decadal Oscillation (PDO); the Southwest US Drought Cycle; the Length of Day (LOD) / Atmospheric Circulation Index; the ThermoHaline Circulation (THC);
El Nino Global-SST ENSO index; the InterTropical Convergence Zone (ITCZ); and the Solar System Influence (CMSS),(see: Global Warming Science (GWS), http://www.appinsys.com/GlobalWarming/, 2011).
In all these series we clearly note the presence of a significant sixty-year quasi-periodicity which have to be considered in all reliable climatic models.
In particular, it is important to include this long-term cycle when we try to extrapolate statistical behaviors for the temperature anomalies.
As well reported in GWS, 2011, the corrected extrapolated linear warming trend, from global temperature anomalies data, when accounting for the 60-yr cycle, is actually about 0.4 degrees per century, as shown by the blue line on the Figure 1, and not 0.74 degrees or more per century
as extrapolated on a shorther time interval (red and green lines in Figure 1). The question now is: how is important the sixty-year
periodicity? The main aim of this note is to give a further contribution to determine if this multidecadal quasi-periodicity can be a possible dominant driver 
of different processes including the climate oscillations. The methodology used here is a multiresolution continuous wavelet analysis which allows
a more detailed investigation of the multiscale periodicity features and their time evolutions.   
    
\begin{figure}
\resizebox{\hsize}{!}{\includegraphics{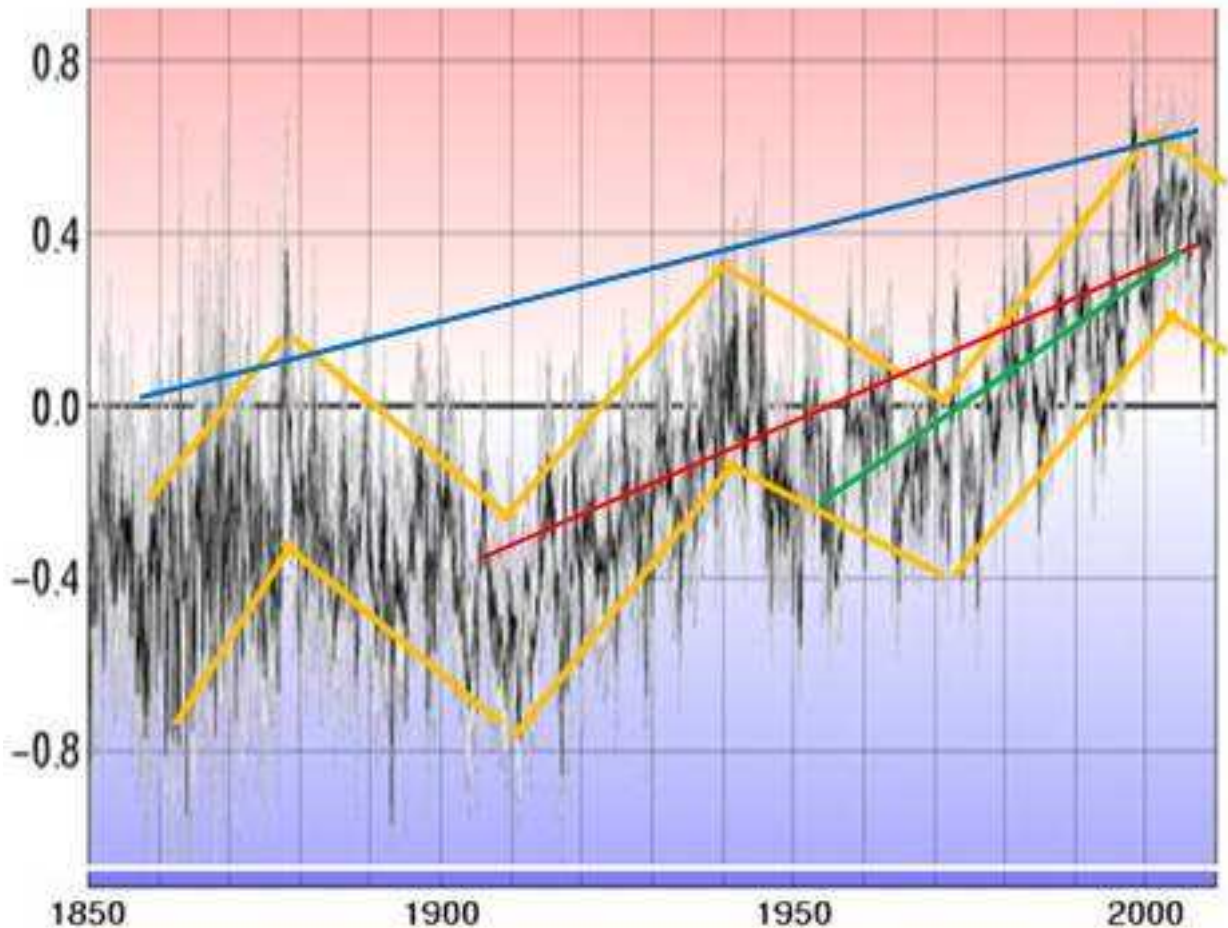}}
 \caption{The global temperature anomalies HadCRUT3, (from: Global Warming Science, (GWS)).
x-axis: Time; y-axis: Temperature Anomaly C.}
 \label{fig1}
\end{figure}

\section{Time series and results}
The selected long-term time series used to investigate the sixty-year periodicity come from different areas: 1) the global monthly temperature
anomalies (period: 1850-March 2011); 2) Global-SST ENSO index (El Nino) (period: 1817-February 2011); 3) Lenght of day (LOD) (period:1876-2003);
4) Solar system CMSS (period: 1850-2099). For details on these time series we refer to the cited bibliography.

The mathematical tool used to analyze the above time series is a multiresolution continuous wavelet method properly designed for both evenly and unevenly sampled data.
The "spatial" resolution in the time-frequency domain is quite high, obtaining a final matrix of 875x875 points. The wavelet basis used here is
a complex Morlet functions set with adjustable parameters. In order to evaluate the statistical significance of the wavelet power spectrum, we compare
the signal spectrum with an equivalent spectrum coming from an autoregressive step-1 Markov process. For futher mathematical details on this wavelet analysis, see: Sello, 2003 and Ranucci and
Sello, 2007.

The following Figure 2 shows the results of the wavelet analysis performed on the first series: the global monthly temperature anomalies data.

Note that the y-axis is shown in logarithmic scale in order to better see all the frequency range of interest involved (here selected from: 1.1 yr to: 105 yr). Black contour lines are boundaries of confidence regions at 95\%.
The color map is related to an arbitrary scale. On the right there is the time-integrated wavelet power showing the main principal peaks (power level is on a logarithmic scale).

\begin{figure}
\resizebox{\hsize}{!}{\includegraphics{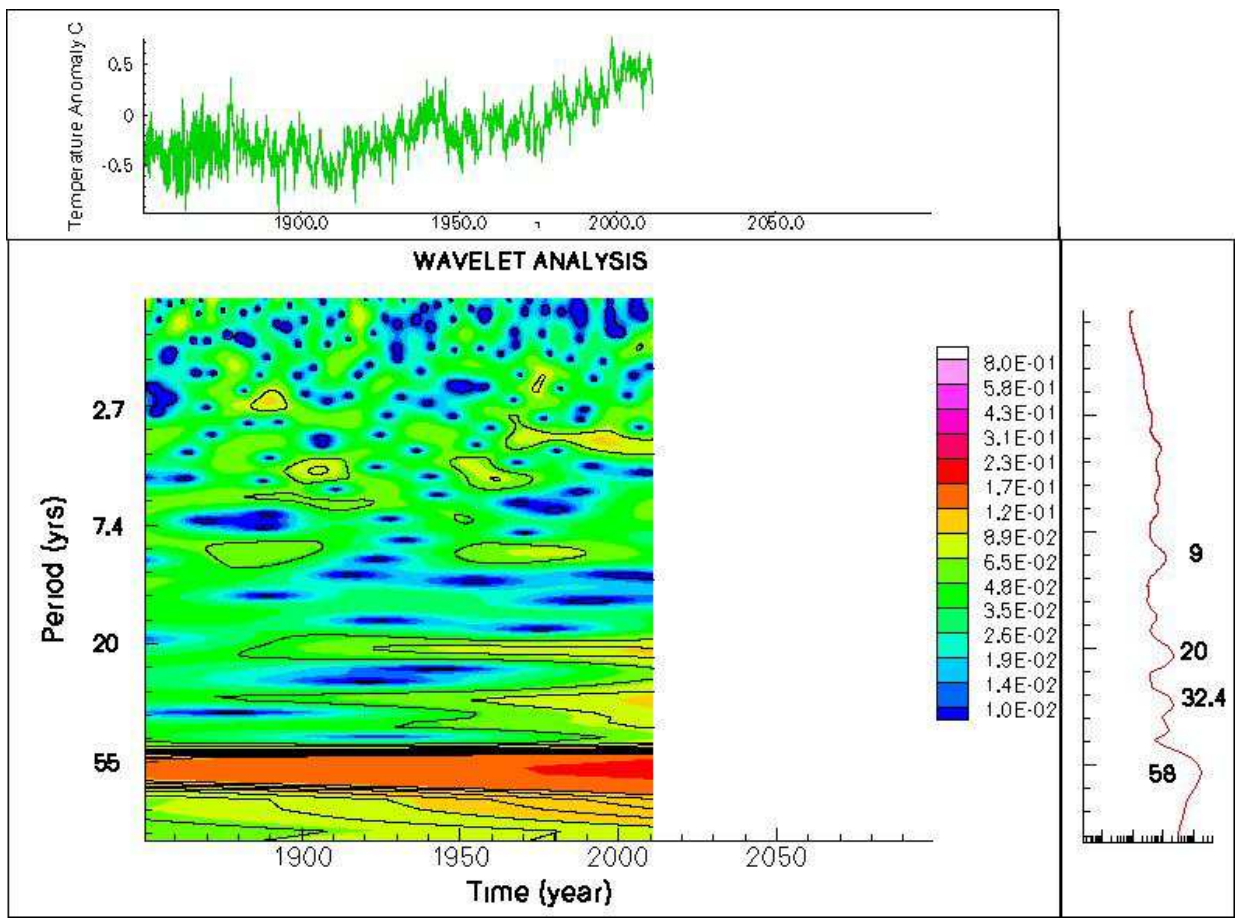}}
 \caption{Wavelet analysis the global monthly temperature anomalies.}
 \label{fig2}
\end{figure}

Our wavelet analysis (WA) confirms the results of Zhen-Shan and Xian, 2007.  The WA map indeed shows that: (1) Temperature anomalies can be completely decomposed into four principal timescales quasi-periodic oscillations including an ENSO-like mode, a near 9 year signal (but not persistent), a 20-year signal (more persistent), a near 32 year more recent signal and a near 60-year signal (both persistent and strong). We confirm also that of the quasi-periodicities detected, the near 60-year timescale oscillation of temperature variations is the most prominent.
Our WA allows us to better define the time-frequency characteristics and the non-linear evolution of the quasi periodicities detected in the signal. 
Thus the conclusion is that all reliable climate models cannot exclude this significant periodicity!

The following Figure 3 shows the wavelet analysis performed on the second series: Global-SST ENSO index (El Nino) data.
Note that the scales of the wavelet are the same as Figure 1. Black contour lines are boundaries of confidence regions at 95\%.

\begin{figure}
\resizebox{\hsize}{!}{\includegraphics{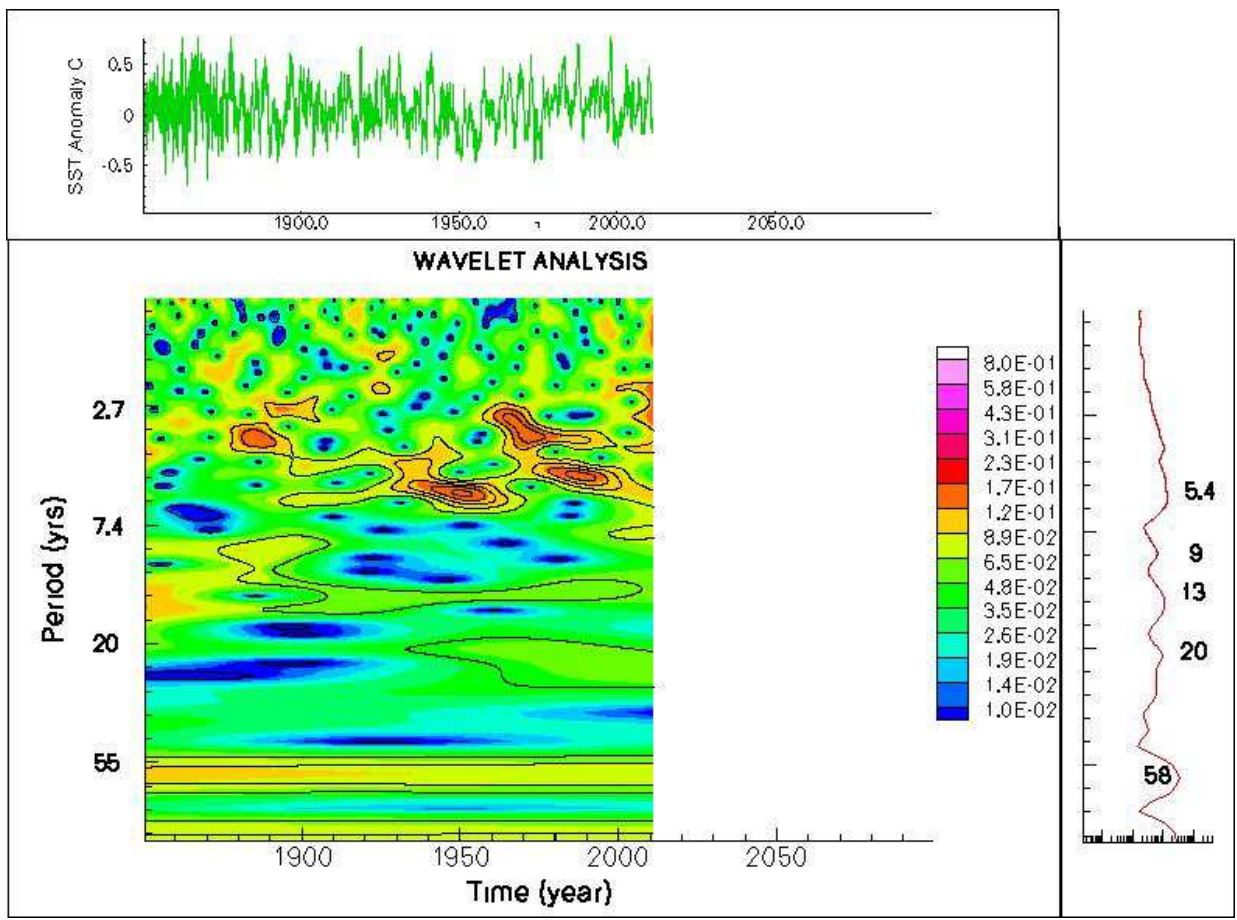}}
 \caption{Wavelet analysis of Global-SST ENSO index (El Nino).}
 \label{fig3}
\end{figure}

Our wavelet analysis clearly shows the presence of a near 60-year persistent periodicity (but more strong in the past: $<$ 1950), and more weak quasi-periodicities at 20, 13 and 9 yrs. Note also the well known strong and time localized ENSO periodicities near 4-5 yrs.

The following Figure 4 shows the wavelet analysis performed on the third series: -LOD data. For more details on this analysis see: Sello, 2011 (arXiv:1103.4924).
Note that the scales of the wavelet are the same as Figure 1, but the contour power levels are here rescaled to the different intensity energy level of the series. 
Black contour lines are boundaries of confidence regions at 95\%.

\begin{figure}
\resizebox{\hsize}{!}{\includegraphics{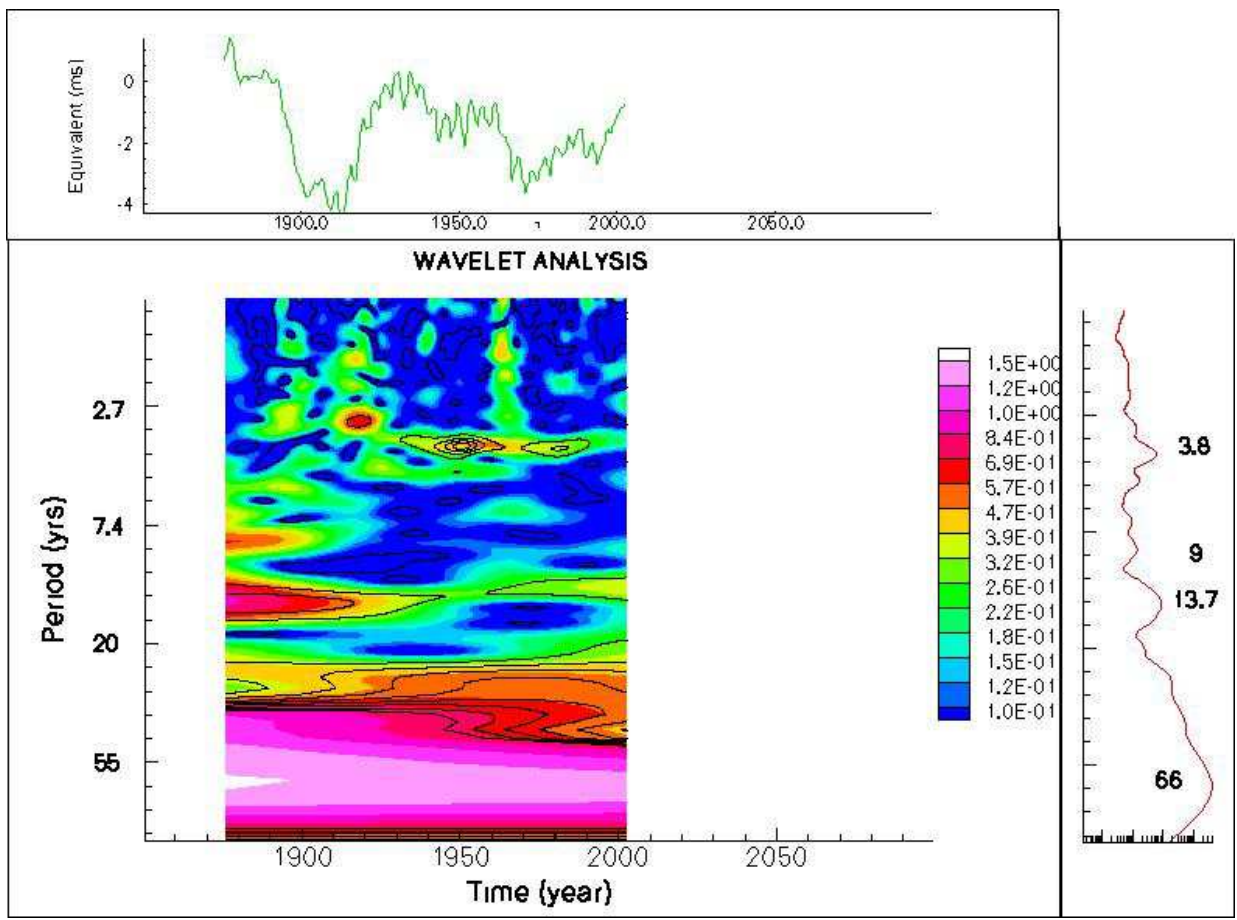}}
 \caption{Wavelet analysis of -LOD data.}
 \label{fig4}
\end{figure}

We confirm again a dominant and time-persistent frequency band in the periodicity range: 50-92 yrs, with a central peak at 66 yr, clearly visible at maximum power level up to 1900. Further, other less persistent frequencies are: 13.7 - 9 - 3.8 yr. Some of these frequencies are well time localized and may be the object of further future investigations.

The following Figure 5 shows the wavelet analysis performed on the fourth series: Solar system CMSS data from NASA Jet Propulsion Laboratory Ephemeris database.
Black contour lines are boundaries of confidence regions at 95\%.

 \begin{figure}
\resizebox{\hsize}{!}{\includegraphics{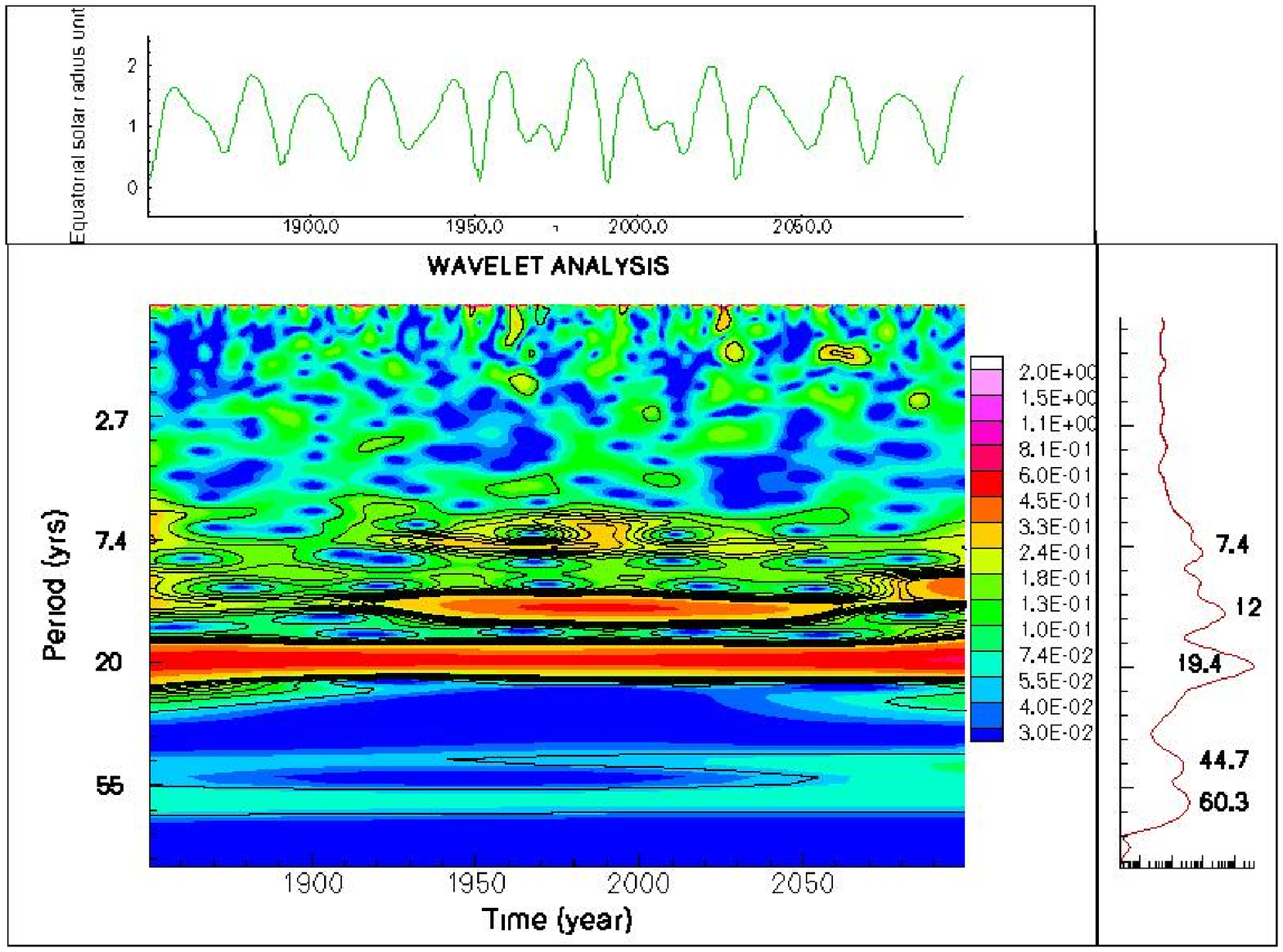}}
 \caption{Wavelet analysis of Solar System CMSS data.}
 \label{fig4}
\end{figure}

As previously suggested by Scafetta, 2010, our WA of CMSS shows very prominent periodicity bands:

A band with central peak at 60.3 yr almost persistent, but almost weak;
A second related band (note that in the future, $>$ 2050, there will be a merge of the two bands) with central peak at 44.7 yr less persistent but always visible.
Note that this periodicity is weak up to 2020, but it will become quite relevant in the following decades, and finally it will be stronger around 2090;
A very strong band with central peak at 19.4 yr and persistent;
A very strong band with central peak at 12 yr less persistent (only recent years) and shifting towards shorter periods (9.8 yr) in the future;
A less persistent and irregular band near 7.4 yr.
A comparison between Figure 2 and Figure 5 shows that, for the temperature anomalies, there is a clear increase of the wavelet power level of the 60-yr periodicity after 1970; whereas the
same behavior is absent for the CMSS. This interval coincides with the time period of an observed acceleration in the increase of the temperature anomalies (see Figure 1).
Even if we cannot here exclude that this feature can be driven by some astronomical term (not considered here), it would be plausible to suggest an anthropic origin.

Thus, it appears quite reasonable the hypothesis of Scafetta that the change in the location of the center of mass of the solar system (CMSS) could be a possible mechanism driving the 60-year cycle (a full cycle of Jupiter / Saturn around the Sun) and that the Jupiter-Saturn conjunction could be the main mechanism for the appearance of a strong and persistent 19.4 yr periodicity. 
Our WA allowed us to better define the time-frequency characteristics and the non-linear evolution of the quasi periodicities detected in the signal. 
For example, it would be interesting to see the effects of the two bands (60.3 yr and 44.7 yr) merging on the driving mechanism that it is suggested to be, at least, partly responsible of multidecadal climate oscillations. 

\section{Conclusions}
The above wavelet analysis results further support the idea, suggested by Scafetta, 2010, and supported by other authors, that the dominant drivers of the long term climate oscillations have a celestial origin. In particular, it is reasonable
to consider the possibility that a significant component of the climate oscillations, with shorter time scale between 1 and 100 years, can be interpreted in astronomical
terms too. Another important evidence, supported by our results, is that all climatic models that aim to describe and predict correctly the past, present and future of global temperatures behavior, cannot
exclude the sixty-year quasi-periodicity which is the dominant and persistent oscillation detected. All statistical extrapolations about the climate time behavior, have to consider
this time scale in order to obtain a correct result. Further investigations, when more data of the above time series will be collected, will allow us to better understand the correlation existing between
the above detected periodicities and the main physical drivers of important planetary processes such as the evolution of climate.   

\section{References}

Global Warming Science (GWS), 2011: http://www.appinsys.com/GlobalWarming/

Komitov, B., 2009: The Sun-climate relationship II: The cosmogenic beryllium and the middle latitude aurora. Bulgarian Astronomical Journal 12, 7590.

Milankovic M. (1941), Canon of insolation and the ice-age problem: (Kanon der Erdbestrahlung und seine Anwendung auf das Eiszeitenproblem), Belgrade, 1941 (Royal Serbian Academy of Mathematical and Natural Sciences, v. 33).

Ogurtsov, M.G., Y.A. Nagovitsyn, G.E. Kocharov, and H. Jungner, 2002: Long-period cycles of the Suns activity recorded in direct solar data and proxies, Solar Phys. 211, 371-394.

Patterson, R. T., A. Prokoph and A. Chang, 2004: Late Holocene sedimentary response to solar and cosmic ray activity influenced climate variability in the NE Pacific. Sedimentary Geology 172, 6784.

Ranucci, G. and Sello, S., 2007: Search for periodicities in the experimental solar neutrino data: A wavelet approach, Phys. Rev. D, 75, 7.

Roberts, P. H., Z. J.Yu, and C. T. Russell, 2007: On the 60-year signal from the core. Geophys. Astrophys. Fluid Dyn., 101, 11-35.
 
Scafetta, N., 2010: Empirical evidence for a celestial origin of the climate oscillations and its implications, arXiv:1005.4639v1. 

Sello, S., 2003: Wavelet entropy and the multi-peaked structure of solar cycle maximum, New Astronomy, Volume 8, Issue 2, p. 105-117.

Sello, S., 2011: On the correlation between air temperature and the core Earth processes: Further investigations using a continuous wavelet analysis, arXiv:1103.4924. 

Yu Z., S. Chang, M. Kumazawa, M. Furumoto, A. Yamamoto, 1983: Presence of periodicity in meteorite falls, National Institute of Polar Research, Memoirs, Special issue (ISSN 0386-0744), 30, 362-366. 

Zhen-Shan, Sun Xian, 2007: Multi-scale analysis of global temperature changes and trend of a drop in temperature in the next 20 years, Meteorology and Atmospheric Physics, 95, 1-2. 

\end{document}